\documentclass[12pt,
    tightenlines,
    onecolumn,
    preprintnumbers,
    amsmath,
    amssymb,
    prd,
    showpacs,
    showkeys
]{revtex4}
\usepackage{colordvi}
\input colordvi
\usepackage{color}

\usepackage{graphicx}%
\usepackage{amsmath}
\usepackage{amssymb}
\usepackage{bm}
\usepackage{enumerate}

\newcommand{\del}{\partial}
\newcommand{\nn}{\nonumber}
\newcommand{\wt}{\widetilde}

\newcommand{\bg}{\textbf{g}}
\newcommand{\mn}{{\mu \nu}}

\newcommand{\cF}{\mathcal{F}}
\newcommand{\cU}{\mathcal{U}}
\newcommand{\tr}{\triangleright}
\newcommand{\btr}{\blacktriangleright}

\newcommand{\cFk}{\mathcal{F_\kappa}}
\newcommand{\bv}{{\bf v}}
\newcommand{\cV}{\mathcal{V}}
\begin{document}
\title{Scalar Field theory in $\kappa$-Minkowski spacetime from twist}

\author{Hyeong-Chan Kim}
\email{hyeongchan@sogang.ac.kr}
\affiliation{Center for Quantum Space Time, Sogang University,
Seoul 121-742, Republic of Korea}
\author{Youngone Lee
}
\email{ youngone@yonsei.ac.kr}
\affiliation{Department of Physics, Daejin University,
Pocheon, 487-711, Republic of Korea}
\author{Chaiho Rim}
\email{rim@chonbuk.ac.kr}
\affiliation{Department of Physics and Research Institute
of Physics and Chemistry, Chonbuk National University, Jeonju 561-756, Korea}
\author{Jae Hyung Yee}
\email{jhyee@yonsei.ac.kr}
\affiliation{Department of Physics, Yonsei University,
Seoul 120-749, Republic of Korea}
\bigskip
\begin{abstract}
\bigskip
Using the twist deformation of $U(igl(4,R))$, the linear part of the diffeomorphism, we define a scalar function and construct a free scalar field theory in four-dimensional $\kappa$-Minkowski spacetime.
The action in momentum space turns out to differ only in integration measure from the commutative theory.
\end{abstract}
\maketitle

\section{Introduction}
Doplicher, Fredenhagen, and Roberts~\cite{Doplicher} showed that, in the presence of gravity, the Heisenberg's uncertainty relation has to be generalized to include the uncertainty between coordinates, which may be reproduced from the noncommutativity of coordinates such as the canonical noncommutativity~\cite{Saxell} or the $\kappa$-Minkowski spacetime~\cite{Majid94,Lukierski95}.
Especially in the $\kappa$-Minkowski spacetime, the position $\hat x^\mu$ satisfies an algebra-like commutational relation,
\begin{eqnarray}\label{kappa}
[x^0, x^i]= \frac{i}{\kappa} x^i,
\end{eqnarray}
with all other commutators vanishing.

The $\kappa$-Minkowski spacetime may arise as an effective low energy description of quantum gravity~\cite{Freidel,Smolin,Amelino}.
Such space first appeared in the investigation of $\kappa$-Poincar\'{e} algebra.
Later, it was related to Doubly Special Relativity (See~\cite{DSR} and references therein) which might have a quantum gravitational origin~\cite{kappa:QG}.

The differential structure of the $\kappa$-Minkowski spacetime has been
constructed in~\cite{kappa-diff} and based on this differential structure,
the scalar field theory has been formulated~\cite{Kosinski,KRY,int-scalar,Freidel,Noether,arzano1,Meljanac}.
It was shown that the differential structure requires that the momentum space
corresponding to the $\kappa$-Minkowski spacetime becomes a de-Sitter section in
five-dimensional flat space.
Various physical aspects of $\kappa$-Minkowski spacetime have been  investigated in Refs.~\cite{KRY} and was extended to $\kappa$-Robertson-Walker spacetime~\cite{kim}.
The Fock space and its symmetries~\cite{amelino,arzano}, $\kappa$-deformed statistics of particles~\cite{das,other2}, and interpretation of the $\kappa$-Minkowski spacetime in terms of exotic oscillator \cite{Ghosh} were also studied.
In addition, the properties of scalar field theory on this spacetime has started being analyzed in depth~\cite{Freidel,Kosinski,Noether,int-scalar}.

Recently, the $\kappa$-Minkowski spacetime is realized in terms of twisting procedure~\cite{lightlike,Bu,other,twist:other2,Ballest}.
This twist approach can be seen as an alternative to the
$\kappa$-like deformation of the quantum
Weyl and conformal algebra~\cite{Ballest},
which is obtained by using the Jordanian
twist~\cite{jordantwist,kulish,Borowiec}.
The light-cone $\kappa$-deformation of Poincar\'e algebra can be given by standard twist  (see eg.~\cite{lightlike}).
The realization for the time-like $\kappa$-deformation was constructed in Ref.~\cite{Bu,other} by embedding an abelian twist in $igl(4,R)$ whose symmetry is bigger than the Poincar\'e, and their differential structure was studied in ~\cite{twist:diff}.
One can also find other approach to the differential structure and twist realization of $\kappa$-Minkowski spacetime by using the Weyl algebra~\cite{twist:other2}.

In this paper, we construct a free scalar field theory by using the twist approach~\cite{Bu,twist:diff}.
In Sec. II, we review the $\kappa$-Minkowski spacetime from twist and then define the $\ast$-product between vectors. We also provide an interesting relation between the generators when acting on coordinates space. %
In Sec. III, we introduce a new action of the generators on function and define a $\star$-product between functions.
In Sec. IV, we find a transformation rule for a scalar function
and then construct an action for a real free scalar field in Sec. V.
We summarize the results and discuss the physical applications in Sec. VI.

\section{Review on the $\kappa$-Minkowski spacetime from twist}
Twisting the Hopf algebra of the universal enveloping algebra of inhomogeneous general linear group is considered in~\cite{Bu,other}.
The group of inhomogeneous linear coordinate transformations is composed of the product of the general linear transformations and the spacetime translations.
The inhomogeneous general linear algebra in (3+1)-dimensional flat spacetime $\bg=igl(4,R)$ is composed of $20$ generators $\{P_a,M^a_{~b}\}$ $(a,b=0,1,2,3)$ where $P_a$ represents the spacetime translation and $M^a_{~b}$ represents the boost, rotations and dilations.
The generators satisfy the commutation relations,
\begin{eqnarray} \label{iglbasis}
\left[P_a, P_b\right] &=&0, \quad
\left[ M^a_{~b}, P_c \right] = i\delta^a_{~c}\cdot P_b, \nn \\
\left[M^a_{~b} ,M^c_{~d}\right] &=&
    i\left(\delta^a_{~d}\cdot M^c_{~b}-\delta^c_{~b}
    \cdot M^a_{~d}\right).
\end{eqnarray}
The universal enveloping Hopf algebra $\cU(\bg,\cdot,\Delta,\varepsilon, S)$ with the counit $\varepsilon$ and antipode $S$ can be constructed starting from the base elements $\{1, P_a, M^a_{~b}\}$ and coproduct $\Delta Y=1\otimes Y+Y\otimes 1$ with $Y\in \{P_a, ~M^a_{~b}\}$.

The infinitesimal transformation by the general linear group is given in terms of $igl(4,R)$ generators:
\begin{eqnarray}\label{action}
\delta_\epsilon S= -i \epsilon^c Y_c \triangleright S\,,
\end{eqnarray}
where $\tr$ denotes an abstract action of $Y$ on vectors, scalar fields, or vector fields.
Consider the action $\tr$ of $Y$ on an algebra $\cV= (V,\cdot)$ where $V\equiv\{ x^a\}\cup\{ k_b\}\cup \{k^c\}$
satisfying
\begin{eqnarray} \label{M:V}
M^a_{~b}\triangleright x^c &=&-i x^a \delta_b^c, \quad M^a_{~b}\triangleright k_c = ik_b\delta^a_c,
   \quad   M^a_{~b}\triangleright k^c = -ik^a\delta_a^c,   \\
P_a \triangleright x^b &=&-i \delta_a^b, \quad\quad  P_a \triangleright k_b = 0, \quad \quad \quad P_a \triangleright k^b=0. \nn
\end{eqnarray}
In the previous paper~\cite{twist:diff}, we considered the vector space $\{x^a\}\cup\{(e_\mu)^b\}$ or equivalently $\{x^a\}\cup\{k^b\}$.
To define a scalar function in this paper we extend the module algebra of $\cU(\bg)$ to include a dual space $\{k_a\}$ to $\{k^a\}$.
Therefore, we generalize the relations in Eqs.~(11) and (12) in Ref.~\cite{twist:diff} to both of the covariant and contravariant vectors.
The action on the product of vectors is given by the coproduct,
\begin{equation}\label{action}
Y\tr (\bv_1\cdot \bv_2) = \cdot (\Delta Y\tr (\bv_1 \otimes \bv_2))=(Y\tr \bv_1)\cdot \bv_2+\bv_1 \cdot (Y\tr \bv_2) ,
\end{equation}
where $\bv_1, \bv_2\in \cV$. If we choose $\bv_1=x^a$ and $\bv_2 =x^b$, this equation provides the well-known Leibnitz rule.

\subsubsection{Abelian twist}
A new Hopf algebra is obtained by twisting a given Hopf algebra.
The new Hopf algebra has the algebra part in common with the original, however, the coproduct is changed by the twist.
A twist $\cF_\kappa$ is a counital 2-cocycle satisfying $(\varepsilon \otimes \mbox{id})\cF_\kappa=1$ and $(1\otimes \cF_\kappa) (\mbox{id}\otimes \Delta)\cF_\kappa =(\cF_\kappa\otimes 1)(\Delta \otimes \mbox{id})\cF_\kappa$.
The new Hopf algebra, $\cU_\kappa(\bg,\cdot,\Delta_\kappa,\varepsilon, S)$ is given by the original counit and antipode ($\varepsilon_{\kappa}= \varepsilon$, $S_\kappa=S$), but with a twisted coproduct:
\begin{equation}
\label{def-coproduct}
\Delta_\kappa(Y)=\cF_\kappa\cdot \Delta Y\cdot \cF_\kappa^{-1}=\sum_{i}Y_{(1)i}\otimes Y_{(2)i}\equiv Y_{(1)} \otimes Y_{(2)} .
\end{equation}

An abelian twist can be constructed by exponentiating two commuting generators such as the momentum operators $P_1$ and $P_2$, which gives the canonical noncommutativity~\cite{Chaichian}.
Other choice of a twist by using two commuting operators $E= P_0$ and $D=\sum_{i=1}^3 M^i_{~i}$~\cite{Bu}
\begin{eqnarray} \label{twist}
{\cal F_\kappa}=
\exp\left[\frac{i}{\kappa}
\Big(\alpha E\otimes D- (1-\alpha) D\otimes E \Big)\right]
\end{eqnarray}
generates the $\kappa$-Minkowski spacetime with the twisted Hopf algebra $\cU_\kappa(\bg)$.
$\alpha$ is a constant representing different ordering of the exponential kernel function in the conventional $\kappa$-Minkowski spacetime formulation.
In this paper, $\alpha=1/2$, which corresponds to the time-symmetric ordering.

For convenience, we explicitly write down the twisted coproduct, ($i,j=1,2,3$)
\begin{eqnarray} \label{iglcoproduct}
\Delta_\kappa(Z) &=& Z \otimes 1+ 1\otimes Z\,, \quad
		Z\in \{ E, D, M^i_{~j}\},  \\
\Delta_\kappa(P_i) &=& P_i\otimes
    e^{E/(2\kappa)}  + e^{-E/(2\kappa)}\otimes P_i, \nn\\
\Delta_\kappa(M^i_{~0}) &=& M^i_{~0}\otimes
    e^{-E/(2\kappa)}  + e^{E/(2\kappa)}\otimes M^i_{~0}, \nn \\
\Delta_\kappa(M^0_{~i}) &=& M^0_{~i}\otimes
    e^{E/(2\kappa)}  + e^{-E/(2\kappa)}\otimes M^0_{~i}+
    \frac{1}{2\kappa}\left(P_i\otimes D e^{E/(2\kappa)}-e^{-E/(2\kappa)} D\otimes P_i\right), \nn \\
\Delta_\kappa(M^0_{~0}) &=& M^0_{~0}\otimes 1  + 1\otimes M^0_{~0}+
    \frac{1}{2\kappa}\left(E\otimes D-D\otimes E\right). \nn
\end{eqnarray}
The spatial (rotational) parts are undeformed and keeps the rotational symmetry.
On the other hand, the boost parts are deformed nontrivially due to the presence of the spatial dilatation term.
It is noted that the twisted Hopf algebra $\cU_\kappa(\bg)$ is different from that of the conventional $\kappa$-Poincar\'e in two aspects.
First, the algebraic part is nothing but those of the un-deformed inhomogeneous general linear group~(\ref{iglbasis}) rather than that of the deformed Poincar\'{e}.
Second, the co-algebra structure is enlarged due to the bigger symmetry $igl(4,R)$ and its co-product is deformed as~(\ref{iglcoproduct}).

\subsubsection{$\ast$-product between vectors}
In this subsection, we study the non-commutative $\ast$-product by using the twisted Hopf algebra.
The twist $\cF_\kappa \in \cU(\bg)$ with new product, $\ast$, given by
\begin{eqnarray}
\bv_1\ast \bv_2\equiv\ast[\bv_1 \otimes \bv_2] &=& \cdot \left[ \cF_\kappa^{-1}
\triangleright (\bv_1 \otimes \bv_2) \right] , \quad \bv_1, \bv_2\in V,
\end{eqnarray}
defines a new associated algebra $\cV_\kappa =(V,\ast)$ as a module algebra of $\cU_\kappa(\bg)$ in the sense that
\begin{eqnarray} \label{Y:ast}
Y\tr( \bv_1\ast \bv_2) =\ast( \Delta_\kappa Y\tr (\bv_1\otimes \bv_2))= \sum (Y_{(1)}\tr \bv_1) \ast (Y_{(2)} \tr \bv_2) .
\end{eqnarray}
Explicitly, the $\ast$-product between $x^a$ and $x^b$ gives the usual $\kappa$-Minkowski relation~(\ref{kappa}) and the $\ast$-product between $x^a$ and $k^b$ leads to the nontrivial commutation relation $[x^0,k^i]= \frac{i}{\kappa}k^i$~\cite{twist:diff}. The $\ast$-product with $k_b$ gives
\begin{eqnarray}\label{x.x}
k_a \ast x^b &=&k_a x^b+\frac{i}{2\kappa} \delta_a^i \delta^b_0 k_i \,,
\\
x^b \ast k_a &=& k_a x^b-\frac{i}{2\kappa} \delta_a^i \delta^b_0 k_i \,,
\nn\\
k_a \ast q_b&=& k_a q_b \,,  k_a \ast \quad q^b = k_a q^b,\nn
\end{eqnarray}
which results in commutation relations
\begin{eqnarray} \label{[x,e]}
[k_a,q_b]_\ast=0,\quad [x^0, k_i]_\ast = -\frac{i}{\kappa} k_i,\quad [x^0, k_0]_\ast=0=[x^i,k_a]_\ast\,.
\end{eqnarray}
Note that any vector having non-vanishing spatial index does not commute with the time coordinate.
It was also shown in Ref.~\cite{twist:diff} that the relation~(\ref{[x,e]}) is related to the $4$-dimensional differential structure of the $\kappa$-Minkowski spacetime from twist.

\subsubsection{Relation between the actions of $M^a_{~b}$ and $P_c$ }
In this subsection, we provide a nontrivial relation satisfied by the two actions of $M^a_{~b}$ and $P_a$ on the vector space of coordinate vector $x^a$:
\begin{eqnarray}\label{M:xn}
M^a_{~b} \tr ( x^{c_1}\ast x^{c_2}\ast\cdots \ast x^{c_n} ) =x^a [P_b    \tr (x^{c_1}\ast x^{c_2}\ast\cdots \ast x^{c_n}  )].
\end{eqnarray}

We prove this by the method of induction.
We use the notation $x^{(n)} = (x^{c_1}\ast x^{c_2}\ast\cdots \ast x^{c_n} )$ for simplicity.
In the case of $x^{(1)}$, it is clear from the definition of $M^a_{~b}$ in Eq.~(\ref{M:V}).
Let us assume that $x^{(n)}$ satisfies Eq.~(\ref{M:xn}).
Then, we can show that $x^{(n+1)}=x^{c_0}\ast x^{(n)}$ also satisfies the relation.
As an illustration we post the proof for the case of $M^0_{~i}$:
\begin{eqnarray} \label{M:xastx}
M^0_{~i} \tr x^{c_0} \ast  x^{(n)} &=& (M^0_{~i}\tr x^{c_0}) \ast (e^{E/(2\kappa)} \tr  x^{(n)})+ (e^{-E/(2\kappa)}\tr x^{c_0})\ast (M^0_{~i}\tr  x^{(n)}) \\
&+&\frac{1}{2\kappa}\left[(P_i\tr x^{c_0})\ast (D e^{E/(2\kappa)}\tr  x^{(n)})
    - (D e^{-E/(2\kappa)}\tr x^{c_0})\ast (P_i \tr  x^{(n)})
    \right] \nn
\end{eqnarray}
where we use the deformed coproduct~(\ref{iglcoproduct}) and the definition~(\ref{Y:ast}) of the action on the $\ast$-product.
The first term in the right-hand-side of Eq.~(\ref{M:xastx}) becomes
\begin{eqnarray*}
-i \delta^{c_0}_i~ x^0\ast  (e^{E/(2\kappa)} \tr  x^{(n)})
 & =& x^0\left[-i\delta^{c_0}_i \ast (e^{E/(2\kappa)} \tr  x^{(n)})\right]-i (\frac{-i}{2\kappa})(-i \delta^{c_0}_i) \ast (De^{E/(2\kappa)}\tr  x^{(n)}) \nn \\
&=& x^0\left[(P_i\tr x^{c_0}) \ast (e^{E/(2\kappa)} \tr  x^{(n)})\right]- \frac{1}{2\kappa}(P_i \tr x^{c_0})\ast (De^{E/(2\kappa)}\tr  x^{(n)}) ,
\end{eqnarray*}
where we use the definition of $\ast$ to replace the $\ast$-product $x^0\ast[\cdots]$ with a normal one $x^0 [\cdots]$, by using $x^0\ast [\cdots]=x^0[\cdots]-\frac{1}{2\kappa}(D\tr [\cdots])$ in the first equality and use the property $P_i \tr x^0=-i \delta_i^{x_0}$ in the second equality.
Similarly, the second term in the right-hand-side of Eq.~(\ref{M:xastx}) becomes
\begin{eqnarray*}
&&(e^{-E/(2\kappa)}\tr x^{c_0})\ast [x^0 (P_{i}\tr  x^{(n)})] \\
&&\quad\quad= (e^{-E/(2\kappa)}\tr x^{c_0})\ast [x^0 \ast (P_{i}\tr  x^{(n)})] +  \frac1{2\kappa}(e^{-E/(2\kappa)}\tr x^{c_0})\ast  (D\tr (P_i \tr  x^{(n)}))   \\
&&\quad\quad = x^0\ast (e^{-E/(2\kappa)}\tr x^{c_0}) \ast (P_i\tr x^{(n)})
        -[x^0,e^{-E/(2\kappa)}\tr x^{c_0}]_\ast\ast( P_i \tr x^{(n)})
        \nn\\
&&\quad\quad~~        +\frac{1}{2\kappa} (e^{-E/(2\kappa)}\tr x^{c_0})\ast (D P_i\tr  x^{(n)}) \\
&&\quad\quad =x^0\left[(e^{-E/(2\kappa)} \tr x^{c_0}) \ast (P_i\tr x^{(n)})\right]
        + \frac{1}{2\kappa} (D e^{-E/(2\kappa)}\tr x^{c_0})    \ast (P_i \tr  x^{(n)}) \nn
\end{eqnarray*}
where we replace the normal product $x^0[\cdots]$ with a $\ast$-product in the first equality, exchange the order of product in the second equality, and then replace the $\ast$-product $x^0\ast[\cdots]$ with a normal product in the last equality.
Adding the above two equations we have
\begin{eqnarray*}
&&(M^0_{~i}\tr x^{c_0}) \ast (e^{E/(2\kappa)} \tr  x^{(n)})+ (e^{-E/(2\kappa)}\tr x^{c_0})\ast (M^0_{~i}\tr  x^{(n)})  \\
 &&= x^0(P_i \tr  x^{(n+1)}) - \frac{1}{2\kappa}\left[ (P_i \tr x^{c_0})\ast (De^{E/(2\kappa)}\tr  x^{(n)})- (D e^{-E/(2\kappa)}\tr x^{c_0}) \ast (P_i \tr  x^{(n)})\right].
\end{eqnarray*}
The two $O(1/\kappa)$ terms exactly cancels the last two terms in Eq.~(\ref{M:xastx}).
In the case of the action of $M^0_{~i}$, one may similarly show by using $x^i\ast [\cdots]=x^i (e^{\frac{E}{2\kappa}} \tr [\cdots])$.
One may similarly demonstrate that the relation~(\ref{M:xn}) is satisfied for other cases too.
This implies that $ x^{(n+1)}$ satisfies Eq.~(\ref{M:xn}).

We emphasize that the action of $M^a_{~b}$ is not equivalent to the action of $x^a (P_b\tr )$ if the target is a tensor composed of  whole module space $\cV_\kappa$ since
\begin{eqnarray}
x^a(P_b\tr k_c \ast x^d) &=&  -i \delta_b^d x^a k_c \neq M^a_{~b}\tr k_c \ast x^d .
\end{eqnarray}
Eq. (\ref{M:xn}) holds only when the target space of $\cU_\kappa(\bg)$ is the coordinates vector space which is a subset of $\cV_\kappa$.

\section{Action on functions and $\star$-product between functions}
We now consider the action of $igl(4,R)$ generators on the algebra of functions $\mathfrak{F}=\{F,m\}$, where $F$ denotes the space of functions and $m$ denotes the ordinary product given by $m[f\otimes g](x)=\cdot [f(x)\otimes g(x)]=f(x) g(x)$, where $f,g\in F$.
For convenience, we introduce a new notation $\btr$ denoting the action $\tr$ of the generators on a function $f$ with the form:
\begin{eqnarray}
M^a_{~b}\btr f(x)&\equiv& \left(M^a_{~b}\triangleright f\right)(x)
     = -i x^a \frac{\partial}{\partial x^b} f(x)\,, \\
P_{~a}\btr f(x) &\equiv&\left(P_a\triangleright f\right)(x)
    = -i \frac{\partial}{\partial x^a} f(x) \,. \nn
\end{eqnarray}
Explicitly, the two actions of $M^a_{~b}$ on a simple function $k(x) = k_a x^a \in F$ give different results,
\begin{equation}\label{M:kx4}
M^a_{~b} \tr (k_a x^a)=0, \quad \quad M^a_{~b}\btr k(x) = -i x^a k_b.
\end{equation}
In the first action $\tr$,  $M^a_{~b}$ acts on both $k_a$ and $x^a$ so that $k_ax^a$ is invariant under $GL(4,R)$ and in the second action $\btr$, it acts only on $x^a$.
On the other hand the two actions of momentum are equivalent:
$$
P_a\btr f(x) = P_a \tr f(x),
$$
since the action of momentum to the vector $k_a$ vanishes.

Given the action $\btr$ of the algebras on $\mathfrak{F}$, we may define an associated algebra $\mathfrak{F}_\kappa =(F, \star)$ as module algebra of $\cU_\kappa(\bg)$ by using the twist $\cF_\kappa$~(\ref{twist}).
The $\star$-product between two functions $f,g\in \mathfrak{F}_\kappa$ is given by twisting the ordinary product $m$ by using the action $\btr$ as
\begin{eqnarray} \label{kmult}
f(x)\star g(x) \equiv m_\kappa[f\otimes g](x)&:=& m\left[\cFk^{-1}\tr
    (f\otimes g)\right](x)  \\
    &=& \cdot\left[\cFk^{-1}\btr(f(x)\otimes g(x))\right]. \nn
\end{eqnarray}
If the functions $f(x)$ and $g(x)$ are ordinary functions without $\star$-product, Eq.~(\ref{kmult}) leads to the conventional $\kappa$-Minkowski star-product,
\begin{eqnarray}  \label{kmoyal}
f(x)\star g(x)&:=&\cdot\left( \exp \left[ \frac{i}{2\kappa}
\Big(D\otimes E -E\otimes D\Big) \right] \btr f(x)\otimes g(x)\right) \\
&=&  \left( \exp \left[ \frac{i}{2\kappa}
\Big(\frac{\del}{\del x_0} y^k\frac{\del}{\del y_k}
-x^k \frac{\del}{\del x_k}\frac{\del}{\del y_0} \Big) \right] (f(x) \cdot g(y)) \right|_{y\to x}. \nn
\end{eqnarray}

One may also try to get a deformed associated algebra by acting the action $\tr$ on $f(x)$ rather than on its functional form $f$, as in Eq.~(\ref{kmult}).
However in this case, the resulting module algebra will be the same as the original one, $\mathfrak{F}$, since any action of generators $M^a_{~b}$ on a scalar makes it vanish.
Thus the star-product $k_a \ast x^a$ defines a scalar function $k$,
$$
k(x)=k_a \ast x^a
$$
due to $k_a \ast x^a=k_a x^a$ and Eq.~(\ref{M:kx4}).
Consider two scalar functions $k(x)=k_a\ast x^a$ and $q(x)=q_a \ast x^a$ so defined.
Explicit calculation shows that the two functions commutes for the $\ast$-product,
$$
k(x) \ast q( x)=q(x)\ast k(x)=k(x) q(x).
$$
This implies that the $\ast$-product between functions reduces to the ordinary product.
However, the $\star$-product does not commute,
\begin{eqnarray} \label{estare}
k(x) q(x)+\frac{i}{2\kappa} \left(k_0 q_i-  k_i q_0 \right)x^i \neq k(x)q(x) .
\end{eqnarray}
From Eqs.~(\ref{x.x}) and (\ref{estare}), one notices that the $\star$-product satisfies
\begin{eqnarray}\label{star:ast}
k( x) \star q( x) =k_a q_b( x^a\ast x^b) \neq q(x)\star k(x).
\end{eqnarray}
This is the reason why we construct the noncommutative star product between scalar functions using $\star$.

Finally, the partial derivative $\partial_a$ is identified with $i P_a$ and its coproduct is defined by $\Delta_\kappa (\partial_a) = i \Delta_\kappa (P_a ) $,
\begin{eqnarray}
\partial_0 \btr (\phi(x)\star \psi(x))
&=& (\partial_0 \btr \phi(x))\star \psi(x)
+ \phi(x)\star (\partial_0 \btr \psi(x)),
\label{Leibnitz} \\
\partial_i \btr(\phi(x)\star \psi(x))
&=&
(\partial_i \btr\phi(x))\star
\Big(e^{\frac E{2\kappa} }\psi(x)\Big)
+ \Big(e^{-\frac E{2\kappa} }\phi(x)\Big)
\star (\partial_i \btr \psi(x)) \,.
\nn
\end{eqnarray}
We have shown in Ref.~\cite{twist:diff} that the differential structure is consistent with the Jacobi identity and the relation $d_\kappa^2=0$.

\section{$\star$-product and vectors}

Up to this point, we have defined the $\star$-product between the scalar functions.
However, it is not yet defined the $\star$-product between a vector and a function.
To find one, we calculate the action of $P_a$ on the product of two scalar functions in two different ways.
First, we calculate it by using the definition of coproduct~(\ref{iglcoproduct}),
\begin{eqnarray}\label{Pbtr:kq}
P_i \btr  \left[k(x) \star q( x)\right]&=&-i k_i \star [e^{E/(2\kappa)} \btr q(x)]
       -i [e^{-E/(2\kappa)}\btr k(x)] \star q_i  \\
&=&-i[ k_i \star q(x)  +k(x)\star q_i] +\frac{1}{2\kappa}\left[k_0 \star q_i-k_i \star q_0\right] , \nn \\
P_0 \btr  \left[k(x) \star q( x)\right]&=&-i[k_0 \star q(x)+k(x) \star q_0 ]\nn.
\end{eqnarray}
Second, we calculate the $\star$-product before acting the momentum operator,
\begin{eqnarray}\label{P:kq2}
P_a \btr  \left[k(x) \star q( x)\right]&=& P_a\btr \left(k(x) q(x)+\frac{i}{2\kappa} \left[k_0 q(x)-  k(x) q_0 \right]\right) \\
  &=& -i[k_a q(x)+k(x) q_a]+\frac{k_0 q_a-k_a q_0}{2\kappa}.
  \nn
\end{eqnarray}
Since the two results should be the same we have the $\star$-product between $k_a$'s and a function:
\begin{eqnarray} \label{k.qx}
k_a\star q(x)=k_a q(x) =q(x)\star k_a, \quad k_a\star q_b=k_a q_b .
\end{eqnarray}
This shows that $k_a$ commutes with the $\star$-product so that
\begin{eqnarray*}
k(x)\star q(x)\star [P_a\btr r(x)]\star\cdots \star s(x)&=&
k(x)\star q(x)\star (-i r_a)\star\cdots \star s(x) \\
&=& -i r_a[ k(x)\star q(x)\star\cdots \star s(x)].
\end{eqnarray*}
The same result holds for the contravariant vector $k^a$.

We can also do the same calculation for $M^a_{~b}$ as in Eqs.~(\ref{Pbtr:kq}) and (\ref{P:kq2}).
For example, we calculate
\begin{eqnarray*}
&&M^0_{~i} \btr \left[k(x) \star q( x)\right] \\
&&=-i [k_i x^0\star q(x)+ k(x)\star x^0q_i]+\frac{k_0\star q_i x^0-x^0k_i \star q_i + x^j k_j\star q_i - k_i \star q_j x^j}{2\kappa} \nn
\end{eqnarray*}
which should be the same as
\begin{eqnarray*}
M^0_{~i} \btr \left[k(x) \star q( x)\right] = x^0\left(-i\left[k_i q(x) + k(x) q_i\right]+
    \frac{1}{2\kappa}\left[ k_0 q_i-  k_iq_0\right]\right).
\end{eqnarray*}
Equating the two equations for all $M^a_{~b}$, we get
\begin{eqnarray} \label{xstare:11}
x^a\star k(x)  &=&  x^a k(x)+\frac{i}{2\kappa}\left(\delta^a_0k_i-\delta^a_i k_0\right)x^i, \\
k(x) \star x^b &=&  k(x) x^b+\frac{i}{2\kappa}\left(k_0\delta^b_i-k_i\delta^b_0 \right)x^i.\nn
\end{eqnarray}

We now provide an independent check of Eqs.~(\ref{k.qx}) and (\ref{xstare:11}).
One may conjecture that $k_a\ast [x^a\star f(x) ]= (k_a\ast x^a)\star f(x)$.
However, one immediately realizes that this conjecture is not valid since Eq.~(\ref{Pbtr:kq}) is different from (\ref{P:kq2}) with this conjecture.
Noting $k_a \ast x^a = k_a x^a$ and Eq.~(\ref{star:ast}), one has to try a weaker form:
\begin{eqnarray} \label{exex2}
k_a [x^a\star q(x) ]= (k_a\ast x^a)\star q(x) =k(x)\star q(x).
\end{eqnarray}
Then, by assuming $x^a\star q(x)=x^a q(x) +\frac{i}{2\kappa}B^a$ in Eq.~(\ref{exex2}),  Eq.~(\ref{estare}) results in Eq.~(\ref{xstare:11}).
In addition, one may have  $k_a\star f(x)= k_a f(x) $ as in Eq.~(\ref{k.qx}) if  one requires
$$
x^a \star [k_a\star f(x)] = k(x)\star f(x)=k_a\star[x^a\star f(x)].
$$

Given the relations~(\ref{k.qx}) and (\ref{xstare:11}), we may show that the actions $M^a_{~b}$ and $P_a$ satisfy
\begin{eqnarray}
 M^a_{~b} \btr \left[k(x) \star q( x)\right]= x^a \left(P_b \btr \left[k(x) \star q( x)\right] \right). \nn
\end{eqnarray}
Similarly, we also get
\begin{eqnarray} \label{M:kn}
M^a_{~b} \btr  \left[k^{1}(x) \star k^{2}( x) \star\cdots  \star k^{n}( x)\right]
    = x^a   \left(P_b \btr \left[k^{1}(x) \star k^{2}( x) \star\cdots  \star k^{n}( x)\right] \right)
\end{eqnarray}
which is consistent with Eq.~(\ref{M:xn}).

In general, one may construct any function as a series of $k(x)$. Therefore, Eq.~(\ref{M:kn}) must be satisfied for all scalar functions.
In this sense, we propose the transformation law of a scalar field $f\in \mathfrak{F}_\kappa$ under a general linear group as
\begin{eqnarray}\label{def:scalarfield}
(M^a_{~b} \tr f)(x) \equiv M^a_{~b} \btr f(x) =  x^a [P_b \btr f(x)] .
\end{eqnarray}

Especially, the time-symmetric exponential function
\begin{eqnarray}\label{[e]}
[e^{ikx}]_s = e^{\frac{i k_0 x^0}{2}} \star e^{i \vec{k} \cdot\vec{x}} \star  e^{\frac{i k_0 x^0}{2}},
\end{eqnarray}
and their product $[e^{i k x}]_s\star [e^{i q x}]_s $ satisfy the property~(\ref{def:scalarfield}).
It is remarked in passing that other ordered exponential functions also satisfy the same property~(\ref{def:scalarfield}).
Note that we have used the $\star$-product to define the ordered exponential function in Eq.~(\ref{[e]}) rather than the $\ast$-product.
This is a crucial difference from the previous work~\cite{twist:diff} in which the ordered product was implemented by using the $\ast$-product and $k_a$ was absent in the module algebra $\cV_\kappa$.

\section{Action of a free scalar field}

In this section, we construct an action of a free scalar field
which is invariant under the action of general linear algebras in the sense of Eqs.~(\ref{iglbasis}) and (\ref{M:V}).

To find a metric we choose $k_a$ and $q_a$ in Eq.~(\ref{estare}) as $(e^\mu)_a$ and $(e^\nu)_b$, the tetrad of the coordinates (Note that the tetrad is independent of the coordinate vector $x^a$ because of Eq.~(\ref{M:V})).
Multiplying the flat Minkowski metric $\eta_{\mu\nu}$ to Eq.~(\ref{estare}) we arrive at
\begin{eqnarray} \label{metric}
 \eta_{\mn}e^\mu(x) \star e^\nu(x)=\eta_{\mu\nu}(e^\mu)_a(e^\nu)_b x^ax^b= g_{ab} x^ax^b.
\end{eqnarray}
$g_{ab}=\eta_{\mu\nu}(e^\mu)_a(e^\nu)_b$ with signature $(-+++)$ transforms under the actions of $igl(4,R)$ generators as
$$
M^a_{~b} \tr g_{cd} = i \left(g_{bd}\, \delta^{a}_c + g_{cb}\,\delta^a_d\right),
 \quad P_a \tr g_{cd} =0 .
$$
Since the RHS of Eq.~(\ref{metric}) is invariant under the general linear transformation, the LHS satisfies the transformation rule,
\begin{equation}
M^a_{~b}  \tr  \eta_{\mn}e^\mu(x) \star e^\nu(x) =0.
\end{equation}

To construct a free scalar field theory in $\kappa$-Minkowski spacetime from twist,
we need to know some useful identities for the time-symmetric exponential function~(\ref{[e]}).
From Eq.~(\ref{k.qx}) we have
\begin{eqnarray} \label{P:eikx}
\partial_a\btr  [e^{i qx}]_s= i q_a [e^{i q_a x^a}]_s.
\end{eqnarray}
(This simple relation is also satisfied by the exponential function of different ordering if one uses different twist parameter $\alpha$).
Eq.~(\ref{P:eikx}) confirms that the exponential function acts as a scalar function following Eq.~(\ref{def:scalarfield}).
In addition, the multiplication of two exponential functions is given by a new scalar exponential function:
\begin{eqnarray} \label{e:star:e}
[e^{ikx}]_s \star [e^{iqx}]_s = [e^{i(k \wt{+} q)x}]_s,
\end{eqnarray}
where $(k\wt{+}q)= (k_0+q_0, k_ie^{\frac{q_0}{2\kappa}}+q_i e^{-\frac{k_0}{2\kappa}})$.
Note that $k\wt{+}q \neq q\wt{+}k$.

In the presence of a non-commutativity we introduce the integration of a scalar function using the property,
$$
\int d^4 x  \sqrt{-g}\star \phi(x) =  \int d^4 x\, \sqrt{-g} \phi(x)
    =   \sqrt{-g}\int d^4 x \, \phi(x)  .
$$
In the first equality, we use  Eq.~(\ref{k.qx}) and in the second equality, we use the fact that the metric tensor $g_{ab}$ is independent of coordinate vector $x^a$.
(If the metric were dependent on coordinate vector, the $\star$-product might be relevant in this calculation.)

We calculate the $\star$-product of two exponential functions to get a $\delta$-function,
$$
\int d^4 x \sqrt{-g}\,[e^{-ikx}]_s\star [e^{iqx}]_s =(2\pi)^4\sqrt{-g} e^{\frac{3q_0}{2\kappa}} \delta^4(k-q) .
$$
Since the left-hand-side is a scalar quantity, the right-hand-side is also a scalar under the transforms in $\cU_\kappa(\bg)$.
Note that the $\delta$-function appears with the normalization factor $e^{\frac{3q_0}{2\kappa}}$.

We define the Fourier transform of a scalar field with the form:
\begin{equation}
\phi(x)=\int d\mu_k\, [e^{i kx}]_s \phi_{k},
\end{equation}
where $d\mu_k$ is an appropriate measure to be determined below.
The inverse Fourier transform must be given by the $\star$-product. We have the consistency condition,
\begin{eqnarray*}
\phi_k&=& \int d^4 x \sqrt{-g} [e^{-i k x}]_s \star \phi(x)
        = \sqrt{-g} \int d\mu_q\,  \phi_{q}(2\pi)^4 e^{\frac{3q_0}{2\kappa}}\,\delta^4(k-q)  ,
\end{eqnarray*}
which determines the measure
\begin{equation} \label{measure}
d\mu_q= \frac{1}{\sqrt{-g}} \frac{d^4q\, e^{\frac{3q_0}{2\kappa}}}{(2\pi)^4} .
\end{equation}
The Hermitian conjugate of $\phi$ becomes
$
\phi^\dagger(x) = \int d\mu_k [e^{i k x}]_s \bar{\phi}_{-k},
$
where $\bar{\phi}$ implies the complex conjugate of $\phi$.
If $\phi(x)$ is a real scalar field, the mode function satisfies $\phi_k = \bar{\phi}_{-k}$.

The integration of the product of two scalar fields is
\begin{eqnarray}\label{phi*phi}
\int d^4x \sqrt{-g}\star \phi(x)\star \psi (x)
&=& \sqrt{-g} \int d^4x\, \phi(x)\star \psi(x)\\
    &=&\frac{1}{ \sqrt{-g} }\int \frac{d^4 k\,d^4q}{(2\pi)^8} e^{\frac{3(k_0+q_0)}{2\kappa}} \phi_k \psi_q \int d^4 x \, [e^{i (k\wt{+}q) x}]_s
        \nn\\
    &=&\int d\mu_k \,\phi_k\psi_{-k}\,. \nn
\end{eqnarray}
One may calculate any number of products of scalar fields in a similar way.

The action for a free scalar field in commutative spacetime with metric $g_{ab}$ is
\begin{eqnarray}
S_{\rm commutative}&=&\frac{1}{2}\int d^4 x \,\sqrt{-g}\left[-g^{ab} (\partial_a \phi(x))
   (\partial_b  \phi(x))- m^2\phi(x) \phi(x)\right].
\end{eqnarray}
This action is invariant under the general linear transformation in $4$-dimensions in the sense of linear diffeomorphism: $M^a_{~b} \tr S =0$ and $P_a \tr S=0$.  
The corresponding action in noncommutative spacetime is obtained by the following modifications: 1) The partial derivative is replaced by the nontrivial action $\partial_a \tr$ on ordered functions.  2) The normal products between functions and vectors are replaced by the $\star$-products.
Since the metric is independent of coordinate vector, the $\star$-product between the metric and  a function is irrelevant.
Thus, the action of a free real scalar field in $\kappa$-Minkowski spacetime from twist is written as
\begin{eqnarray}\label{S}
S&=&\frac{1}{2}\int d^4 x \,\sqrt{-g}\left[-g^{ab} (\partial_a \btr \phi(x))\star
   (\partial_b \btr \phi(x))- m^2\phi(x)\star \phi(x)\right].
\end{eqnarray}
Note the way how $P_a\btr [e^{ikx}]_s$ transforms under the action $M^a_{~b}\btr$:
\begin{eqnarray*}
M^a_{~b}\btr( P_c\btr [e^{ikx}]_s) = [M^a_{~b}, P_c]\btr [e^{ikx}]_s + P_c \btr(x^a (P_b\btr   [e^{ikx}]_s))
    = k_b k_c x^a [e^{ikx}]_s.
\end{eqnarray*}
Therefore,  $P_c\btr [e^{ikx}]_s$ transforms the same way as that of a scalar function under $M^a_{~b}$.

By using Eq.~(\ref{phi*phi}), the action~(\ref{S}) can be expressed in terms of Fourier modes in a quite simple form,
\begin{eqnarray}
S=-\frac{1}{2}\int d\mu_k \,\phi_k(k^2+m^2)\phi_{-k},
\end{eqnarray}
where $k^2 = g^{ab}k_a k_b$.
The onshell condition now becomes $k^2 = -m^2$, which is the same as that of the commutative space field theory.
Therefore, the action of a free scalar field in momentum space is the same as that of the commutative free scalar field theory except for the measure.
This is reasonable since the twist deformation changes the multi-particle sector only and the free action describes the one particle motion.

The equation of motion of a free scalar field in $\kappa$-Minkowski spacetime from twist can be obtained by varying the action~(\ref{S}) with respect to $\phi(x)$ to be
$$
-g^{ab}\partial_a \partial_b \btr \phi(x) +m^2 \phi(x) =0 ,
$$
where we use the partial integration with the help of the modified Leibnitz rule~(\ref{Leibnitz}),
\begin{eqnarray*} \label{partialInt}
&&\int d^4 x\sqrt{-g}[\partial_i \btr \phi(x)] \star \psi(x) \\
 &&~~=
 \int d^4 x\sqrt{-g}\,\partial_i\btr [\phi(x) \star e^{-E/(2\kappa)} \psi(x)]
 -\int d^4 x\sqrt{-g}\phi(x) \star [\partial_i\btr \psi(x)]  \nn \\
 &&~~ = -\int d^4 x\sqrt{-g}\phi(x) \star [\partial_i\btr \psi(x)], \nn
\end{eqnarray*}
where the first term of the second line vanishes which contributes only from the boundary at infinity.

\section{Summary and Discussions}
We have constructed a free scalar field theory on $\kappa$-Minkowski spacetime from twist deformation of inhomogeneous general linear group, which is invariant in the sense of diffeomorphism.
There are two crucial steps.
First, we extend the definition of the $\ast$-product to arbitrary vectors including coordinate vectors and coordinate independent covariant and contravariant vectors.
Second, we introduce the action ($\btr$) of $igl(4,R)$ generators on a function to distinguish it from the action ($\tr$) on vectors.
Given the action ($\btr$) on functions, we define the $\star$-product between functions in terms of the action.

This can be summarized as Eq.~(\ref{xstare:11}): {\it In the calculation of the $\star$-product, the $igl(4,R)$ algebras act only on the coordinate vectors and not on non-coordinate vectors}.
We proposed that all non-commutative product between functions should be given by $\star$-product not by $\ast$ since the $\ast$-product between two scalar functions is trivial.
We also showed that a scalar field in a noncommutative spacetime transforms under the action of $igl(4,R)$ as
$$
M^a_{~b} \btr f(x) = x^a [P_b \btr f(x)] .
$$
Once these conceptual issues are resolved, the construction of a free scalar field theory in $\kappa$-Minkowski spacetime from twist is straightforward.

In the twist deformation, the one particle sector of the scalar field is the same as that of the commutative field theory except for the measure factor.
Therefore, the pole structure of the propagator is the same as that of the commutative theory.
Especially, there are no complex poles unlike in the conventional $\kappa$-Poincar\'{e} formulation.
The true nature of spacetime noncommutativity appears in the multi-particle sector of the theory.
Therefore, it is interesting to study the interacting field theory and statistical effects.
We may add an interaction term such as $\frac{\lambda}6 \int d^4 x \sqrt{-g}\phi(x)\star \phi(x)\star \phi(x)$ to the free action~(\ref{S}).
The term becomes
\begin{eqnarray*}
\frac{\lambda}{6} \int d^4 x\sqrt{-g}\phi(x)\star \phi(x)\star \phi(x)
& =& \frac{\lambda}6 \sqrt{-g}\int d\mu_p d\mu_q d\mu_rd^4 x \phi_p \phi_q\phi_r [e^{i(p\wt{+}q\wt{+}r)x}]_s \\
 &=& \frac{\lambda}6 \int d\mu_p d\mu_q\, \phi_p \phi_q\phi_{-(p\wt{+}q)}.
\end{eqnarray*}
It appears that the energy momentum conservation for bare vertex changes nontrivially as in the  conventional scalar field theory in $\kappa$-Minkowski spacetime~\cite{Kosinski}.
Nonetheless, the propagator in our formulation is different from the conventional one, and
it would be interesting to study how the loop corrections differ from the conventional one.



\begin{acknowledgments}
This work was supported by the Korea Science and Engineering Foundation (KOSEF) grant funded by the Korea Government(MEST)(R01-2008-000-21026-0 (R)) and through the Center for Quantum Spacetime(CQUeST) of Sogang University with grant number R11-2005-021(K\&R) and by the Korea Research Foundation Grant funded by the Korean Government(MOEHRD)(KRF-2008-314-C00063(K\&L)).
\end{acknowledgments} \vspace{3cm}


\begin{thebibliography}{10}
\bibitem{Doplicher}
S. Doplicher, K. Fredenhagen, J. E. Roberts,
``The Quantum Structure of Spacetime at the Planck Scale and Quantum Fields,"
Commun. Math. Phys. {\bf 172}, 187 (1995),
[arXiv:hep-th/0303037].

\bibitem{Saxell}
S. Saxell,
``On general properties of Lorentz invariant formulation of noncommutative quantum field theory,"
Phys. Lett. {\bf B666}, 486 (2008),
[arXiv:0804.3341 hep-th].

\bibitem{Majid94}
S. Majid and H. Ruegg,
``Bicrossproduct structure of $\kappa$-Poincare group and noncommutative geometry",
Phys. Lett. {\bf B334}, 348 (1994)
[arXiv:hep-th/9405107].

\bibitem{Lukierski95}
J. Lukierski, H. Ruegg, and W. J. Zakrzewski,
``Classical quantum mechanics of free $\kappa$-relativistic systems",
Ann. of Phys. {\bf 243}, 90 (1995)
[arXiv:hep-th/0405273].

\bibitem{Freidel}
L. Freidel, J. Kowalski-Glikman, and S. Nowak,
``From noncommutative kappa-Minkowski to Minkowski space-time",
Phys. Lett.  {\bf B648}, 70 (2007)
[arXiv:hep-th/0612170].

\bibitem{Smolin}
L. Smolin,
``An invitation to loop quantum gravity",
[arXiv:hep-th/0408048].

\bibitem{Amelino}
G. Amelino-Camelia,
``Doubly Special Relativity",
Nature {\bf 418}, 34 (2002)
[arXiv:gr-qc/0207049].


\bibitem{DSR}
J. Kowalski-Glikman,
``Introduction to doubly special relativity",
Lect. Notes Phys. {\bf 669}, 131 (2005)
[arXiv:hep-th/0405273].

\bibitem{kappa:QG}
G. Amelino-Camelia, L. Smolin, and A. Starodubtse,
``Quantum symmetry, the cosmological constant and Planck scale phenomenology",
Class. Quant. Grav. {\bf 21}, 3095 (2004)
[arXiv:hep-th/0306134];
L. Freidel, J. Kowalski-Glikman and L. Smolin,
``2+1 gravity and doubly special relativity",
Phys. Rev. {\bf D69}, 044001 (2004)
[arXiv:hep-th/0307085].

\bibitem{kappa-diff}
A.\ Sitarz,
``Noncommutative Differential Calculus on the Kappa-Minkowski Space,"
Phys.\ Lett.\ {\bf B349}, 42 (1995);
C.\ Gonera, P.\ Kosi\'{n}ski, and P.\ Ma\'{s}lanka,
``Differential calculi on quantum Minkowski space,"
J.\ Math.\ Phys.\ {\bf 37}, 5820 (1996)
[arXiv:q-alg/9602007].

\bibitem{Kosinski}
P.\ Kosi\'{n}ski, J.\ Lukierski, and P.\ Ma\'{s}lanka,
``Local D=4 Field Theory on $\kappa$--Deformed Minkowski Space,"
Phys.\ Rev.\ {\bf D62}, 025004 (2000)
[arXiv:hep-th/9902037].

\bibitem{Noether}
A. Agostini, G. Amelino-Camelia, M. Arzano, A. Marciano, and R. A. Tacchi,
``Generalizing the Noether theorem for Hopf-algebra spacetime symmetries",
[arXiv:hep-th/0607221];
M. Arzano and A. Marciano,
``Symplectic geometry and Noether charges for Hopf algebra spacetime symmetries",
Phys. Rev. {\bf D75}, 081701 (2007) [arXiv:hep-th/0701268];
L. Friedel, J. Kowalski-Glikman, and S. Nowak,
``Field theory on $\kappa$-Minkowski spacetime revisited: Noether charges and breaking of Lorentz symmetry",
[arXiv:0706.3658].

\bibitem{int-scalar}
C.\ Rim, ``Interacting scalar field theory in $\kappa$-Minkowski spacetime"
[arXiv:0802.3793].

\bibitem{arzano1}
G. Amelino-Camelia and M. Arzano,
`` Coproduct and star product in field theories on Lie-algebra non-commutative space-times,"
Phys. Rev. {\bf D 65}, 084044 (2002)
[arXiv:hep-th/0105120].

\bibitem{Meljanac}
S. Meljanac, A. Samsarov, M. Stoji\'c, and K.S. Gupta,
``Kappa-Minkowski space-time and the star product realizations,"
Eur.Phys.J. {\bf C53}, 295 (2008)
[arXiv:hep-th/0705.2471].

\bibitem{KRY}
H.-C.\ Kim. J.\ H.\ Yee, and C.\ Rim,
``$\kappa-$Minkowski spacetime and a uniformly accelerating observer,"
Phys. Rev. {\bf D75}, 045017 (2007)
[arXiv:hep-th/0701054].
H.-C.\ Kim, C.\ Rim, and J.\ H.\ Yee,
``Blackbody radiation in $\kappa$-Minkowski spacetime,"
Phys.\ Rev.\ {\bf D76}, 105012 (2007)
[arXiv:0705.4628].
H.-C.\ Kim, C.\ Rim, and J.\ H.\ Yee,
``Casimir energy of a spherical shell in $\kappa$-Minkowski spacetime,"
J. Korean Phys. Soc. {\bf 53}, 1826 (2008)
[arXiv:07105633];
``Symmetric ordering effect on Casimir energy in $\kappa$-Minkowski spacetime''
[arXiv:0803.2333].

\bibitem{kim}
H.-C.\ Kim, J.\ H.\ Yee, and C.\ Rim,
``Density fluctuations in $\kappa$-deformed inflationary universe,"
Phys.\ Rev.\ {\bf D72}, 103523 (2005)
[arXiv:gr-qc/0506122].


\bibitem{amelino}
A.\ Agostini, G.\ Amelino-Camelia, and F.\ D'Andrea,
`` Hopf-algebra description of noncommutative-spacetime symmetries,"
Int.\ J.\ Mod.\ Phys.\ {\bf A19}, 5187  (2004)
[hep-th/0306013];
A.\ Agostini,
`` Fields and symmetries in $\kappa$-Minkowski noncommutative spacetime,"
[hep-th/0312305] (Ph. D. Thesis).

\bibitem{arzano}
M. Arzano and A. Marcian\`{o},
``Fock space, quantum fields and kappa-Poincare symmetries,"
Phys. Rev. {\bf D76}, 125005 (2007)
[arXiv:hep-th/0707.1329].

\bibitem{das}
M.\ Daszkiewicz, J.\ Lukierski, and M.\ Woronowicz,
``$\kappa$-Deformed Statistics and Classical Fourmomentum Addition Law,"
Mod.\ Phys.\ Lett.\ {\bf A23}, 9 (2008)
[hep-th/0703200].

\bibitem{other2}
C.\ A.\ S.\ Young and R.\ Zegers,
``Covariant particle exchange for $\kappa$-deformed theories in 1+1 dimensions,"
Nucl. Phys. {\bf B797}, 537 (2008)
[arXiv:0711.2206].

\bibitem{Ghosh}
S. Ghosh and P. Pal,
``$\kappa$-Minkowski Spacetime Through Exotic Oscillator,"
Phys. Lett. {\bf B618}, 243 (2005)
[arXiv:hep-th/0502192].

\bibitem{lightlike}
A. Borowiec, J. Lukierski, and V. N. Tolstoy,
Eur. Phys. J. {\bf C44}, 139 (2005)
[arXiv:hep-th/0412131].
%

\bibitem{Bu}
J.-G.\ Bu, H.-C.\ Kim, Y.\ Lee, C.\ H.\ Vac, and J.\ H.\ Yee,
``Kappa-deformed Spacetime From Twist",
Phys.\ Lett.\ {\bf B665}, 95 (2008)
[arXiv:hep-th/0611175].

\bibitem{other}
T.\ R.\ Govindarajan, K.\ S.\ Gupta, E.\ Harikumar,
S.\ Meljanac, and D.\ Meljanac,
``Twisted Statistics in kappa-Minkowski Spacetime",
Phys.\ Rev.\ {\bf D77}, 105010 (2008)
[arXiv:0802.1576].

\bibitem{twist:other2}
S.\ Meljanac and S.\ Kresic-Juric,
``Generalized kappa-deformed spaces, star-products, and their realizations",
J.\ Phys.\ A: Math.\ Theor.\ {\bf 41}, 235203 (2008)
[arXiv:hep-th/0804.3072];
``Noncommutative differential forms on the $kappa$-deformed space,"
[arXiv:hep-th/0812.4572].

\bibitem{Ballest}
A. Ballesteros, N. R. Bruno, F. J. Herranz,
``A non-commutative Minkowskian spacetime from a quantum AdS algebra",
Phys. Lett. {\bf B574}, 276 (2003)
[arXiv:hep-th/0306089].

\bibitem{jordantwist}
O. V. Ogievetsky, in {\it Proc. Winter School Geometry
and Physics, Zidkov, Suppl. Rendiconti cir. Math. Palermo, Series II - N 37}, 185 (1993);
preprint MPI-Ph/92-99, Munich (1992).

\bibitem{kulish}
P. P. Kulish, V. D. Lyakhovsky, and A. I. Mudrov,
``Extended jordanian twists for Lie algebras",
[arXiv:math/9806014];
P. P. Kulish, V. D. Lyakhovsky, and M. A. D. Olmo,
``Chains of twists for classical Lie algebras",
[arXiv:math/9908061].

\bibitem{Borowiec}
A. Borowiec and A. Pachol,
``kappa-Minkowski spacetime as the result of Jordanian twist deformation,"
[arXiv:math-ph/0812.0576].

\bibitem{twist:diff}
H.-C. Kim, Y. Lee, C. Rim, and J. H. Yee,
``Differential structure on the $\kappa$-Minkowski spacetime from twist",
[arXiv:hep-th/0808.2866].

\bibitem{Chaichian}
M. Chaichian, P. Kulish, K. Nishijima, A. Tureanu,
``On a Lorentz-Invariant Interpretation of Noncommutative Space-Time and Its Implications on Noncommutative QFT,"
Phys. Lett. {\bf B604}, 98 (2004),
[arXiv:hep-th/0408069].

\end{thebibliography}
\end{document}